# Graded extension of SO(2,1) Lie algebra and the search for exact solutions of Dirac equation by point canonical transformations

A. D. Alhaidari

Physics Department, King Fahd University of Petroleum & Minerals, Box 5047,
Dhahran 31261, Saudi Arabia
E-mail: haidari@mailaps.org

**Abstract:** SO(2,1) is the symmetry algebra for a class of three-parameter problems that includes the oscillator, Coulomb and Mörse potentials as well as other problems at zero energy. All of the potentials in this class can be mapped into the oscillator potential by point canonical transformations. We call this class the "oscillator class". A nontrivial graded extension of SO(2,1) is defined and its realization by two-dimensional matrices of differential operators acting in spinor space is given. It turns out that this graded algebra is the supersymmetry algebra for a class of relativistic potentials that includes the Dirac-Oscillator, Dirac-Coulomb and Dirac-Mörse potentials. This class is, in fact, the relativistic extension of the oscillator class. A new point canonical transformation, which is compatible with the relativistic problem, is formulated. It maps all of these relativistic potentials into the Dirac-Oscillator potential.

## I. INTRODUCTION

In the potential algebra approach [1,2], the dynamical symmetry of the physical problem is exploited by studying the representations of its spectrum generating algebra. Realization of the generators of such algebra by differential operators facilitates study of the properties of the wave equation and its solutions. Examples of these are: SO(2,1) algebra for solving some three-parameter problems such as the harmonic oscillator [1,2]; the conformal group algebra SO(4,2) in the study of the extended symmetry of the Coulomb and Kepler problems [3]; SO(3,2) algebra in the investigation of the singular representations of Dirac supermultiplet (the "singletons") [4]. Potentials with dynamical symmetries associated with any one of these algebras are grouped into classes. All potentials in a given class, along with their corresponding solutions (energy spectrum and wavefunctions), can be mapped into one another by "point canonical transformation (PCT)" [5]. Henceforth, only one problem (the "reference problem") in a given class needs to be solved to obtain solutions of all other problems in the class. PCT maintains the functional form of the problem (i.e. shape invariance of the potential). In other words, it leaves the form of the wave equation invariant. Therefore, it gives a correspondence map among the potential parameters, angular momentum, and energy of the two problems (the new and reference problem). Using the parameter substitution map and the bound states spectrum of the reference problem one can easily and directly obtain the spectra of all other potentials in the class. Moreover, the eigenstates wavefunctions are obtained by simple transformations of those of the reference problem. An alternative approach is to start with a problem whose exact solution is well established then apply to it PCTs that preserve the structure of the wave equation resulting in new exactly solvable problems that belong to the class of the original (reference) problem. Thus the reference problem acts like a seed for generating new exact solutions. This approach is suitable in the search for solutions of a given wave equation by making an exhaustive study of all PCTs that maintain shape invariance of the given wave equation.

In nonrelativistic quantum mechanics, this development was carried out over the years by many authors where several classes of shape invariant potentials being accounted for and tabulated [6]. It was also extended to new classes of conditionally exactly and quasi exactly solvable potentials [7,8] where all or, respectively, part of the energy spectrum is known. However, no such scheme exists in relativistic quantum mechanics. In this article, which is part of our program of searching for exact solutions to the Dirac equation [9-11] and the fourth in the series, this development will be carried out. A nontrivial graded extension of SO(2,1) Lie algebra, which is the relevant potential algebra for our current problem, is defined. Realization of this superalgebra by two-dimensional matrices of differential operators acting in spinor space will be given. The linear span of this superalgebra gives the radial Dirac operator, which carries a representation of this supersymmetry. An associated extension of PCT is formulated and used to obtain new exact solutions of Dirac equation starting from a reference problem, which is taken to be the Dirac-Oscillator problem [12].

SO(2,1) Lie algebra is the dynamical symmetry algebra for many three parameter problems in nonrelativistic quantum mechanics [1,2]. The class of potentials associated with this algebra includes the oscillator, Coulomb, and Mörse potentials, as well as, a large number of power-law potentials at zero energy [11]. The wavefunction solutions of these problems are all written in terms of the confluent hypergeometric functions. All of these problems can be mapped into the oscillator problem by PCTs. For this reason we name this class the "oscillator class". In the Appendix we give a brief overview of the discrete representations and operator realization of SO(2,1) algebra, as well as, the PCTs that generate the oscillator class. In section II the graded extension of SO(2,1) algebra is defined and its realization by two dimensional matrices of differential operators acting in the two-component spinor apace is given. The linear span of this superalgebra is identified with the canonical form of the radial Dirac operator. The resulting relativistic wave equation is solved for the Dirac-Oscillator reference potential. The energy spectrum and spinor wave functions for the Dirac-Oscillator are reproduced. In section III an extended PCT (XPCT), which is compatible with the relativistic problem and the two-component radial Dirac wave equation, is formulated. In section IV this new XPCT is applied to the Dirac-Oscillator problem [12] to obtain exact solutions of Dirac equation that belong to this class. These include the Dirac-Coulomb and Dirac-Mörse problems [9] in addition to relativistic problems at rest energies with power-law potentials [11].

## II. GRADED EXTENSION OF SO(2,1) ALGEBRA

Using gauge invariance of QED, it has been shown [9,10] that the 2-component radial Dirac equation for a charged spinor in spherically symmetric electromagnetic 4-potential can generally be written, in atomic units ($m = e = \hbar = 1$) and taking the speed of light $c = \alpha^{-1}$, as

$$\begin{pmatrix} 1+\alpha^2 \mathcal{V}(r) & \alpha\left(\dfrac{\kappa}{r}+\mathcal{W}(r)-\dfrac{d}{dr}\right) \\ \alpha\left(\dfrac{\kappa}{r}+\mathcal{W}(r)+\dfrac{d}{dr}\right) & -1+\alpha^2 \mathcal{V}(r) \end{pmatrix} \begin{pmatrix} \phi(r) \\ \theta(r) \end{pmatrix} = \varepsilon \begin{pmatrix} \phi(r) \\ \theta(r) \end{pmatrix} \qquad (2.1)$$

where $\alpha$ is the fine structure constant, $\varepsilon$ is the relativistic energy and $\kappa$ is the spin-orbit coupling parameter defined as $\kappa = \pm(j + \tfrac{1}{2})$ for $l = j \pm \tfrac{1}{2}$. The real radial functions $\mathcal{V}(r)$ and $\mathcal{W}(r)$ are the even and odd components of the relativistic potential, respectively. $\mathcal{W}(r)$ is a gauge potential that does not contribute to the magnetic field. However, fixing this gauge degree of freedom by taking $\mathcal{W} = 0$ is not the best choice. An alternative and proper "gauge fixing condition", which is much more fruitful, will be imposed as a constraint in equation (2.2) below. Now, equation (2.1) gives two coupled first order differential equations for the two radial spinor components. By eliminating the lower component we obtain a second order differential equation for the upper. The resulting equation may turn out to be not Schrödinger-like, i.e. it may contain first order derivatives. We apply a global unitary transformation, $U$, parameterized by a real angle parameter $\rho$, that eliminates the first order derivatives:

$$U = \exp\left(\tfrac{i}{2}\sigma_2 \rho\right) = \begin{pmatrix} \cos(\rho/2) & \sin(\rho/2) \\ -\sin(\rho/2) & \cos(\rho/2) \end{pmatrix} \;,\; \text{where } \sigma_2 = \begin{pmatrix} 0 & -i \\ i & 0 \end{pmatrix}$$

The stated requirement gives a constraint ("gauge fixing condition") that relates the even and odd components of the relativistic potential as follows:

$$\mathcal{V}(r) = \dfrac{S}{\alpha}\left[\mathcal{W}(r) + \dfrac{\kappa}{r}\right] \qquad (2.2)$$

It also maps the radial Dirac equation (2.1) into

$$\begin{pmatrix} C + 2\alpha S\left(\mathcal{W}+\dfrac{\kappa}{r}\right) & \alpha\left(-\dfrac{S}{\alpha}+C\mathcal{W}+\dfrac{C\kappa}{r}-\dfrac{d}{dr}\right) \\ \alpha\left(-\dfrac{S}{\alpha}+C\mathcal{W}+\dfrac{C\kappa}{r}+\dfrac{d}{dr}\right) & -C \end{pmatrix} \begin{pmatrix} \phi(r) \\ \theta(r) \end{pmatrix} = \varepsilon \begin{pmatrix} \phi(r) \\ \theta(r) \end{pmatrix} \qquad (2.3)$$

where $S \equiv \sin(\rho)$ and $C \equiv \cos(\rho) = \pm\sqrt{1-S^2}$. Equation (2.3) gives the lower spinor component in terms of the upper as follows:

$$\theta = \dfrac{\alpha}{C+\varepsilon}\left[-\dfrac{S}{\alpha} + C\left(\mathcal{W}+\dfrac{\kappa}{r}\right) + \dfrac{d}{dr}\right]\phi \qquad (2.4)$$

While, the 2$^{nd}$ order radial Schrödinger-like differential equation for the upper component reads

$$\left[-\dfrac{d^2}{dr^2} + \dfrac{C\kappa(C\kappa+1)}{r^2} + \dfrac{2\kappa S\varepsilon/\alpha}{r} + \right.$$
$$\left. + C^2\mathcal{W}^2 + \dfrac{2S\varepsilon}{\alpha}\mathcal{W} - C\dfrac{d\mathcal{W}}{dr} + 2\kappa C^2 \dfrac{\mathcal{W}}{r} - \dfrac{\varepsilon^2-1}{\alpha^2}\right]\phi = 0 \qquad (2.5)$$

Next, we will establish that the canonical form of the Dirac operator in (2.3) (when $\rho = 0$) can be obtained as realization of a superalgebra which is a nontrivial graded extension of SO(2,1) Lie algebra. We start by defining this four dimensional superalgebra $G$ which has two odd elements $L_\pm$ and two even elements $L_0$, $L_3$. It satisfies the following commutation / anticommutation relations:

$$[L_3, L_\pm] = \pm L_\pm$$
$$\{L_+, L_-\} = L_0 \quad (2.6)$$
$$[L_0, L_3] = [L_0, L_\pm] = 0$$

where $L_\pm^\dagger = L_\mp$ which implies hermiticity of the even operators, that is $L_0^\dagger = L_0$ and $L_3^\dagger = L_3$. These relations also show that $L_0$ forms the center of this superalgebra since it commutes with all of its elements. We construct a realization of the generators of this graded Lie algebra as 2×2 matrices of differential operators acting in a two-dimensional $L^2(0,\infty)$ space (the two-component spinor space). The odd operators, which are first order in the derivatives, are the raising and lowering operators in this 2-dimensional spinor space:

$$L_+ = \begin{pmatrix} 0 & G(x) - \dfrac{d}{dx} \\ 0 & 0 \end{pmatrix}, \quad L_- = \begin{pmatrix} 0 & 0 \\ G(x) + \dfrac{d}{dx} & 0 \end{pmatrix}, \quad L_3 = \frac{1}{2}\begin{pmatrix} 1 & 0 \\ 0 & -1 \end{pmatrix}$$

Using the anticommutation relation in (2.6) we obtain

$$L_0 = \begin{pmatrix} -\dfrac{d^2}{dx^2} + G^2 - G' & 0 \\ 0 & -\dfrac{d^2}{dx^2} + G^2 + G' \end{pmatrix} \quad (2.7)$$

where $G(x)$ is a real function and $G' \equiv dG/dx$. It is to be noted that the odd operators are first order in the derivatives whereas the even operators are zero and second order. If an operator $Q \in G$ (i.e. carries a representation of this supersymmetry), then $Q$ can be expanded as a linear combination of these four basis operators as follows:

$$Q = \lambda_+ L_+ + \lambda_- L_- + \lambda_3 L_3 + \lambda_0 L_0$$

Requiring that this operator be hermitian and first order differential gives

$$\lambda_+^* = \lambda_-, \lambda_3^* = \lambda_3, \lambda_0 = 0$$

which yields

$$Q = \begin{pmatrix} 1 & \alpha\left(G - \dfrac{d}{dx}\right) \\ \alpha\left(G + \dfrac{d}{dx}\right) & -1 \end{pmatrix}$$

where $\alpha \equiv 2\lambda_+/\lambda_3$ and $Q$ has been renormalized with the nonzero "mass factor" $\lambda_3/2$. Note that $Q$ is equivalent to the canonical form of Dirac Hamiltonian in equation (2.1) with $\mathcal{V}(r) = 0$ [equivalently, equation (2.3) with $\rho = 0$] and $G(r) = \kappa/r + \mathcal{W}(r)$. The two-dimensional spinor eigenvector of $Q$ satisfies the following eigenvalue equation

$$(Q - \varepsilon)\begin{pmatrix} f_+(x) \\ f_-(x) \end{pmatrix} = 0$$

where $\varepsilon$ is real. Since $L_0$ is the center of the superalgebra (i.e. a constant scalar, say $2E$), then using equation (2.7), the components of this eigenvector also satisfy

$$\left(-\frac{d^2}{dx^2} + G^2 \mp G'\right) f_{\pm}(x) = 2E f_{\pm}(x) \tag{2.8}$$

This is a one dimensional Schrödinger equation with superpartner potentials
$$V_{\pm}(x) = G^2 \pm G'$$
It is known from supersymmetric quantum mechanics [13] that the zero energy eigenvalue belongs only to the potential $V_-$ while all other eigenvalues are degenerate. Note also the equivalence of equation (2.8) for $V_-$, with $E = (\varepsilon^2 - 1)/2\alpha^2$ and $G = \kappa/r + \mathcal{W}$, to equation (2.5) with $\rho = 0$.

The Dirac-Oscillator reference problem is defined with the following choice of angle parameter and odd potential component:
$$\rho = 0 \quad \text{and} \quad \mathcal{W}(r) = \lambda^2 r$$
where $\lambda$ is the oscillator strength parameter. With this choice equation (2.5) gives the following 2$^{nd}$ order Schrödinger-like differential equation for the upper spinor component:
$$\left[-\frac{d^2}{dr^2} + \frac{\kappa(\kappa+1)}{r^2} + \lambda^4 r^2 + (2\kappa-1)\lambda^2 - \frac{\varepsilon^2 - 1}{\alpha^2}\right]\phi(r) = 0$$
Comparing this equation with the differential equation (A.5) for the nonrelativistic oscillator in the Appendix (with $l \equiv 2\gamma + \frac{1}{2}$) we conclude that $\kappa = l$ or $\kappa = -l-1$ and that the energy spectrum is
$$\varepsilon_n = \sqrt{1 + 2\alpha^2 \lambda^2 (2n + l + \kappa + 1)} \quad ; n = 0,1,2,\ldots$$
Moreover, the normalized solution of this equation is obtained from (A.6) by correspondence as
$$\phi_n(r) = a_n (\lambda r)^{\kappa+1} e^{-\lambda^2 r^2/2} L_n^{\kappa+1/2}(\lambda^2 r^2) \tag{2.9}$$
where the normalization constant $a_n = \sqrt{2\lambda \, \Gamma(n+1)/\Gamma(n+\kappa+3/2)}$. Equation (2.4) gives the lower spinor component in terms of the upper. Using the differential and recursion properties of the Laguerre polynomials [14], we can write it explicitly as
$$\theta_n(r) = \frac{2\alpha\lambda a_n}{\varepsilon_n + 1}(n + \kappa + 1/2)(\lambda r)^{\kappa} e^{-\lambda^2 r^2/2} L_n^{\kappa-1/2}(\lambda^2 r^2) \tag{2.10}$$
In the following section we define an extension of PCT (XPCT) that is compatible with the relativistic wave equation (2.3). It will then be used to obtain exact solutions of Dirac equation for relativistic potentials that belong to the same class as that of the Dirac-Oscillator.

III. THE EXTENDED POINT CANONICAL TRANSFORMATION

We define an extension of PCT that preserves the structure of Dirac wave equation (2.3). It maps a reference problem into new exact solutions of Dirac equation that belong to the same class as the reference problem and carries a representation of the supersymmetry defined in section II. To make the development clear, we choose a notation where the configuration space coordinates for the reference problem is $x$ while that for the new problem is $r$. Moreover, common parameters and variables are distinguished with the caret symbol for those belonging to the reference problem.

Let ψ stands for the radial two-component spinor wavefunction whose upper and lower components are $\phi$ and $\theta$, respectively. Moreover, let $H$ stands for the radial Dirac wave operator in equation (2.3). We define XPCT as the transformation

$$r = q(x) \quad , \quad \phi(r) = g(x)\hat{\phi}(x) \quad \text{and} \quad \theta(r) = h(x)\hat{\theta}(x) \tag{3.1}$$

that leaves the form of the relativistic wave equation (2.3) invariant. $q(x)$, $g(x)$, and $h(x)$ are real differentiable functions. XPCT maps the wave equation (2.3) into the following

$$(H - \varepsilon)\psi = R\left[R^{-1}(H - \varepsilon)R\right]\hat{\psi} = 0 \tag{3.2}$$

where the transformation matrix $R = \begin{pmatrix} g & 0 \\ 0 & h \end{pmatrix}$. Shape invariance of XPCT in (3.1) requires that equation (3.2) be identified with the reference wave equation $(\hat{H} - \hat{\varepsilon})\hat{\psi} = 0$. That is, we can write

$$\hat{H} - \hat{\varepsilon} = D\left[R^{-1}(H - \varepsilon)R\right]$$

where $D$ is an arbitrary 2×2 nonsingular real matrix (i.e. its determinant does not vanish). Substituting from (2.3) and (3.1) into the above equation, we find that hermiticity of the transformed wave operator requires that $h'/h = -g'/g$ giving $h(x) = \xi/g(x)$ where $\xi$ is a real nonzero parameter. Moreover, the same hermiticity requirement puts the matrix $D$ into the following form

$$D = \begin{pmatrix} q'g^2/\xi & 0 \\ 0 & \xi q'/g^2 \end{pmatrix}$$

The nonsingular requirement for $D$ necessitates that $q' \neq 0$. Putting all of that together we obtain the transformed wave operator $(H - \varepsilon)$ in terms of the transformation functions $q(x)$ and $g(x)$ and their derivatives with respect to the coordinate $x$. Shape invariance of XPCT, which in this case requires that the resulting 2$^{nd}$ order differential equation for the transformed upper spinor component be Schrödinger-like, results in that $g(x) = \sqrt{dq/dx}$. Finally, we obtain the following transformed Dirac wave equation

$$\begin{pmatrix} \dfrac{q'^2}{\xi}\left[C + 2\alpha S\left(\mathcal{W} + \dfrac{\kappa}{q}\right) - \varepsilon\right] & \alpha\left[-q'\dfrac{S}{\alpha} + Cq'\left(\mathcal{W} + \dfrac{\kappa}{q}\right) + \dfrac{1}{2}\dfrac{q''}{q'} - \dfrac{d}{dx}\right] \\ \alpha\left[-q'\dfrac{S}{\alpha} + Cq'\left(\mathcal{W} + \dfrac{\kappa}{q}\right) + \dfrac{1}{2}\dfrac{q''}{q'} + \dfrac{d}{dx}\right] & -\xi(C + \varepsilon) \end{pmatrix} \begin{pmatrix} \hat{\phi} \\ \hat{\theta} \end{pmatrix} = 0$$

This has to be identified with the wave equation (2.3) for the reference problem giving $\xi = (\hat{\varepsilon} + \hat{C})/(\varepsilon + C)$ and two important relations. The <u>first</u> one reads:

$$-\dfrac{S}{\alpha} + C\left[\mathcal{W}(r) + \dfrac{\kappa}{r}\right] = \dfrac{1}{q'}\left\{-\dfrac{\hat{S}}{\alpha} + \hat{C}\left[\hat{\mathcal{W}}(x) + \dfrac{\hat{\kappa}}{x}\right] - \dfrac{1}{2}\dfrac{q''}{q'}\right\} \tag{3.3}$$

which is an equality relation modulo a constant due to the derivative term $d/dx$. It gives the new spin-orbit coupling κ in terms of the reference problem parameters. Moreover, it also results in an expression for the odd component of the new relativistic potential, $\mathcal{W}(r)$, in terms of the reference problem and the transformation function $q$. The <u>second</u> relation, which is obtained by equating the determinants of the two wave operators (equivalently, the corresponding 2$^{nd}$ order differential equations), reads

$$F(r) = (q')^{-2} \left[ \hat{F}(x) + \frac{1}{2}\frac{q'''}{q'} - \frac{3}{4}\left(\frac{q''}{q'}\right)^2 \right]$$ (3.4)

$$\equiv \frac{C\kappa(C\kappa+1)}{r^2} + \frac{2\kappa S\varepsilon/\alpha}{r} + C^2\mathcal{W}^2 + \frac{2S\varepsilon}{\alpha}\mathcal{W} - C\frac{d\mathcal{W}}{dr} + 2\kappa C^2\frac{\mathcal{W}}{r} - \frac{\varepsilon^2-1}{\alpha^2}$$

where the expression for $\hat{F}(x)$ is identical to the second line in the above relation except that $r \to x$ and all parameters and variables refer to the reference problem. Substituting the results obtained from (3.3), including $\mathcal{W}(r)$, into the second relation (3.4) gives the energy spectrum for the new problem. Therefore, we end up with a correspondence map among all parameters of the two problems. Using this map and the transformation relation (3.1) we obtain, as well, the two component spinor wavefunctions for the new problem from those of the reference problem.

## IV. THE DIRAC-OSCILLATOR CLASS OF POTENTIALS

Using XPCT developed in the previous section and the Dirac-Oscillator as reference problem, we generate new exact solutions of Dirac equation. Taking the Dirac-Oscillator problem as reference simplifies manipulation because $\hat{\rho} = 0$ (i.e. $\hat{S} = 0$ and $\hat{C} = 1$). Moreover, the function $\hat{F}(x)$ in equation (3.4) is obtained from the results in the previous section simply as

$$\hat{F}(x) = \frac{\hat{\kappa}(\hat{\kappa}+1)}{x^2} + \lambda^4 x^2 - 2\lambda^2\left(2n + |\hat{\kappa} + \tfrac{1}{2}| + 1\right)$$

We will consider various choices of XPCT function, $q(x)$, and obtain the corresponding relativistic potentials and their solutions (energy spectrum and spinor wavefunctions). Here, we will not attempt to make an exhaustive study of all such XPCTs, however, we will certainly reproduce all known exact solutions of Dirac equation in this class including the most recent results regarding power-law potentials at zero kinetic (rest) energy [11].

### A. The Dirac-Coulomb Problem:

Taking $q(x) = x^2$ and substituting in relation (3.3) give the following:

$\mathcal{W}(r) = 0$

$\kappa = (\hat{\kappa} - \tfrac{1}{2})/2C$

Using the gauge fixing condition (2.2), we obtain the following form for the even component of the relativistic potential

$$\mathcal{V}(r) = \frac{S\kappa/\alpha}{r} \equiv \frac{Z}{r}$$

This is the Coulomb potential where $Z$ is the particle charge number. Therefore, the sine parameter can be written as $S = \alpha Z/\kappa$, while the cosine parameter is

$$C = \pm\sqrt{1 - S^2} = \pm\sqrt{1 - (\alpha Z/\kappa)^2} \equiv \sigma/\kappa$$

With these results, equation (2.5) gives the following 2$^{nd}$ order Schrödinger-like differential equation for the upper spinor component

$$\left[ -\frac{d^2}{dr^2} + \frac{\sigma(\sigma+1)}{r^2} + 2\varepsilon\frac{Z}{r} - \frac{\varepsilon^2-1}{\alpha^2} \right]\phi = 0$$

which shows that $\sigma$ stands for the relativistic angular momentum. Now, the second correspondence relation (3.4) results in the following energy spectrum

$$\varepsilon_n = \left[1 + \left(\frac{\alpha Z}{n + \frac{1}{2} + |\sigma + \frac{1}{2}|}\right)^2\right]^{-1/2} \quad ; n = 0, 1, 2, \ldots$$

The spinor wavefunctions are obtained from (2.9) and (2.10) using the transformation (3.1) and the parameter map giving

$$\phi_n(r) = \sqrt{q'}\hat{\phi}(x) = a_n(\mu_n r)^{\sigma+1} e^{-\mu_n r/2} L_n^{2\sigma+1}(\mu_n r)$$

$$\theta_n(r) = \frac{\xi}{\sqrt{q'}}\hat{\theta}(x) = \frac{\alpha \mu_n a_n}{\varepsilon_n + \sigma/\kappa}(2\sigma + n + 1)(\mu_n r)^{\sigma} e^{-\mu_n r/2} L_n^{2\sigma}(\mu_n r)$$

where $\mu_n = -2Z\varepsilon_n/(n + \frac{1}{2} + |\sigma + \frac{1}{2}|)$ and $a_n$ is the normalization constant.

**B. The Dirac-Mörse Problem:**

We choose $q(x) = -(2/\tau)\ln x$, where $\tau$ is a real positive parameter. Substitution in relation (3.3) gives the following:

$\kappa = 0$

$\mathcal{W}(r) = -(\tau\lambda^2/2C)e^{-\tau r}$

Equation (2.5), results in the following 2nd order differential equation for the upper spinor component

$$\left[-\frac{d^2}{dr^2} + \left(\frac{\tau\lambda^2}{2}\right)^2 e^{-2\tau r} - \frac{\tau\lambda^2}{2}\left(\tau + \frac{2T}{\alpha}\varepsilon\right)e^{-\tau r} - \frac{\varepsilon^2 - 1}{\alpha^2}\right]\phi(r) = 0$$

where $T = S/C$. The second correspondence relation (3.4) gives the following equation for the energy spectrum

$$\varepsilon^2 + \left[\alpha\tau\left(T\varepsilon/\alpha\tau - n\right)\right]^2 = 1$$

which can be written as $\cos^2\varphi + \sin^2\varphi = 1$, where $\varepsilon = \cos\varphi$. Therefore, we need to solve the following equation for the angle $\varphi_n$:

$$\frac{T}{\alpha\tau}\cos\varphi_n - \frac{1}{\alpha\tau}\sin\varphi_n = n \quad ; n = 0, 1, \ldots, n_{max} \leq \sqrt{1 + T^2}/\alpha\tau$$

which can be rewritten as:

$$\sin(\rho)\cos(\varphi_n) - \cos(\rho)\sin(\varphi_n) = n\alpha\tau\cos(\rho)$$

Its solution is $\varphi_n = \rho - \sin^{-1}(n\alpha\tau\cos\rho)$ resulting in the following elegant analytic formula for the energy spectrum:

$$\varepsilon_n = \cos\{\rho - \sin^{-1}[n\alpha\tau\cos(\rho)]\} \tag{4.1}$$

This is an alternative but equivalent result to that which was obtained in reference [9]. The spinor wavefunctions are also obtained using the transformation in (3.1) and the Dirac-Oscillator wavefunctions in (2.9) and (2.10) giving

$$\phi_n(r) = \sqrt{q'}\hat{\phi}(x) = a_n(\lambda^2 e^{-\tau r})^{\nu_n} \exp(-\tfrac{\lambda^2}{2}e^{-\tau r})L_n^{2\nu_n}(\lambda^2 e^{-\tau r})$$

$$\theta_n(r) = \frac{\xi}{\sqrt{q'}}\hat{\theta}(x) = \frac{\alpha\tau a_n}{\varepsilon_n + \frac{1}{\sqrt{1+T^2}}}(2\nu_n + n)(\lambda^2 e^{-\tau r})^{\nu_n}\exp(-\tfrac{\lambda^2}{2}e^{-\tau r})L_n^{2\nu_n-1}(\lambda^2 e^{-\tau r})$$

where $\nu_n = T\varepsilon_n/\alpha\tau - n$ and $a_n$ is the normalization constant.

**C. Zero Energy Problems:**

We consider now the transformation function $q(x) = x^{2\mu+1}$, where $\mu$ is a real parameter such that $\mu \neq 0, \pm\frac{1}{2}$. The three dismissed values of $\mu$ correspond to the Dirac-Oscillator, Dirac-Coulomb, and Dirac-Mörse problems, respectively. Substitution in relation (3.3) gives the following:

$$\mathcal{W}(r) = \frac{\lambda^2/(2\mu+1)C}{r^{(\mu-\frac{1}{2})/(\mu+\frac{1}{2})}}$$

$$\kappa = (\hat{\kappa} - \mu)/(2\mu+1)C$$

On the other hand, the second correspondence relation (3.4) gives $S = 0$ and $\varepsilon = 1$. Compatibility of these results yields ONLY one of two possibilities

(1) $n = 0$ , $\kappa = l$ , $\mu < -\frac{1}{2}$,  OR

(2) $n = 0$ , $\kappa = -l-1$ , $\mu > -\frac{1}{2}$

We write $\mu = -\frac{1}{2} + \beta^{-1}$, where $\beta$ is real, finite and $1 \neq \beta \neq 2$. Thus, we can summarize the physical parameters of the new zero energy system and the odd component of its relativistic potential as follows:

$$\varepsilon = 1, \quad n = 0$$

$$\mathcal{W}(r) = \frac{\beta\lambda^2/2}{r^{1-\beta}}$$

$$\kappa = l, \beta < 0 \quad \text{OR} \quad \kappa = -l-1, \beta > 0$$

Equation (2.5), gives the following 2$^{nd}$ order differential equation for the upper spinor component:

$$\left[-\frac{d^2}{dr^2} + \frac{\kappa(\kappa+1)}{r^2} + \frac{(\beta\lambda^2/2)^2}{r^{2(1-\beta)}} + (2\kappa - \beta + 1)\frac{\beta\lambda^2/2}{r^{2-\beta}}\right]\phi(r) = 0$$

The spinor wavefunctions obtained using the transformation (3.1) and the Dirac-Oscillator spinor components in (2.9) and (2.10) (with $n = 0$ and $\hat{\kappa} \to -\hat{\kappa} - 1$), are:

$$\phi_\kappa(r) = \sqrt{q'}\hat{\phi}_0(x) = a_\kappa \left(\lambda^{2/\beta} r\right)^{-\kappa} e^{-\frac{1}{2}\lambda^2 r^\beta}$$

$$\theta_\kappa(r) = \frac{\xi}{\sqrt{q'}}\hat{\theta}_0(x) = -\alpha\lambda^{2/\beta}(\kappa + \frac{1}{2})a_\kappa \left(\lambda^{2/\beta} r\right)^{-\kappa-1} e^{-\frac{1}{2}\lambda^2 r^\beta}$$

These results are to be compared with those found in reference [11].

There is another class of exactly solvable relativistic potentials for the Dirac equation [10]. It includes Dirac-Rosen-Mörse I & II, Dirac-Eckart, Dirac-Pöschl-Teller, and Dirac-Scarf potentials. In principle, a similar treatment can be carried out where the reference problem is taken to be associated with any suitable one of these potentials, e.g. the Dirac-Rosen-Mörse.

APPENDIX : SO(2,1) POTENTIAL ALGEBRA AND THE OSCILLATOR CLASS

In this Appendix we give a brief overview of the SO(2,1) Lie algebra, its discrete representations, and operator realization that is suitable for the solution of physical problems which are relevant to our case. SO(2,1) is a three dimensional Lie algebra whose generators satisfy the commutation relations

$$[L_3, L_\pm] = \pm L_\pm$$

$$[L_+, L_-] = -L_3$$

where $L_3^\dagger = L_3$ and $L_\pm^\dagger = L_\mp$. It is an algebra of rank one and has one Casimir invariant operator. This operator, which commutes with $L_3$ and $L_\pm$, is $L^2 = L_3(L_3 \pm 1) - 2L_\mp L_\pm$. Among the four operators $L^2$, $L_3$ and $L_\pm$ there is a maximum of two commuting, one of which is $L^2$. To obtain the discrete representation, we choose the compact operator $L_3$ to commute with $L^2$ rather than the non-compact $L_\pm$. Therefore, $L_3$ shares the same eigenvectors with $L^2$. Elements of this representation are labeled with two parameters corresponding to the eigenvalues of $L^2$ and $L_3$. In fact there are three discrete unitary representations of SO(2,1) [1]. Two of them are bounded at one end; one with a lower bound and the other with an upper bound. The third is not bounded. Physically, we are interested in the one that is bounded by a ground state from below. It is parameterized by a real constant $\gamma \geq -1/2$ and denoted as $D^+(\gamma)$. The action of the operators of the algebra on the basis $|\gamma, n\rangle$ is as follows:

$$L^2 |\gamma, n\rangle = \gamma(\gamma + 1)|\gamma, n\rangle$$
$$L_3 |\gamma, n\rangle = (\gamma + n + 1)|\gamma, n\rangle$$
$$L_+ |\gamma, n\rangle = \frac{1}{\sqrt{2}}\sqrt{(n+1)(n+2\gamma+2)}|\gamma, n+1\rangle \quad ; n = 0, 1, 2, \ldots \quad (A.1)$$
$$L_- |\gamma, n\rangle = \frac{1}{\sqrt{2}}\sqrt{n(n+2\gamma+1)}|\gamma, n-1\rangle$$

Realization of the generators of SO(2,1) algebra in terms of differential operators in one variable is of great importance since it is intended to solve the physically interesting second order differential wave equations of the form:

$$\left[\frac{d^2}{dx^2} + f(x)\right]\Phi(x) = 0 \quad (A.2)$$

where

$$f(x) = -\frac{l(l+1)}{x^2} - 2V(x) + 2E \quad (A.3)$$

$l$ is the angular momentum quantum number, $E$ is the energy, and $V(x)$ is a real potential function. The most general form of the three generators $L_3$ and $L_\pm$ whose linear combination gives the second order differential operator in (A.2) are:

$$L_3 = \frac{d^2}{dx^2} + b_3(x)\frac{d}{dx} + a_3(x)$$
$$L_\pm = \frac{1}{\sqrt{2}}\left[\frac{d^2}{dx^2} + b_\pm(x)\frac{d}{dx} + a_\pm(x)\right]$$

Hermiticity and the fact that in the space of $L^2(0,\infty)$ functions, $\overrightarrow{(d/dx)}^\dagger = -\overrightarrow{(d/dx)}$ give $b_- = -b_+$, $a_- = a_+ - db_+/dx$, and $b_3 = 0$. Moreover, applying the commutation relations and performing some manipulations, we arrive at the following:

$$L_3 = \frac{d^2}{dx^2} + \frac{\eta}{x^2} - \frac{x^2}{16}$$
$$L_\pm = \frac{1}{\sqrt{2}}\left[\frac{d^2}{dx^2} + \frac{\eta}{x^2} + \frac{x^2}{16} \pm \frac{1}{2}\left(x\frac{d}{dx} + \frac{1}{2}\right)\right]$$

where $\eta$ is a real constant parameters. As a result of this realization, one finds that

$$L^2 = -\frac{1}{4}\left(\eta + \frac{3}{4}\right) \equiv \gamma(\gamma+1)$$

Thus giving $\eta = -4\gamma(\gamma+1) - 3/4$, and $\eta \leq \frac{1}{4}$. The Hamiltonian of a system whose spectrum is generated by the representation of SO(2,1), must be an element in the linear vector space spanned by its generators. That is we can expand such a Hamiltonian, $H$, as a linear combination of these generators and as follows:

$$-2H = \sqrt{2}\tau_+ L_+ + \sqrt{2}\tau_- L_- + \tau_3 L_3$$

where $\tau_\pm^* = \tau_\mp$, $\tau_3^* = \tau_3$ are constant parameters. Using the realization of $L_3$ and $L_\pm$ obtained above, we can write this as

$$-2H = (\tau_+ + \tau_- + \tau_3)\frac{d^2}{dx^2} + \frac{1}{2}(\tau_+ - \tau_-)x\frac{d}{dx} + (\tau_+ + \tau_- + \tau_3)\frac{\eta}{x^2}$$
$$+ (\tau_+ + \tau_- - \tau_3)\frac{x^2}{16} + \frac{1}{4}(\tau_+ - \tau_-)$$

Therefore, $\tau_+ + \tau_- + \tau_3 = 1$. Moreover, to obtain a Schrödinger-like equation, the first order derivative has to be eliminated which requires that $\tau_+ = \tau_- = (1-\tau_3)/2$ giving:

$$-2H = (1-\tau_3)L_1 + \tau_3 L_3 = \frac{d^2}{dx^2} + \frac{\eta}{x^2} + \frac{1-2\tau_3}{16}x^2$$

where $L_\pm = (L_1 \pm iL_2)/\sqrt{2}$. Schrödinger wave equation is the eigenvalue equation $(H-E)\Phi = 0$ which can now be written as

$$\left[(1-\tau_3)L_1 + \tau_3 L_3 - \tau_0\right]\Phi = \left(\frac{d^2}{dx^2} + \frac{\eta}{x^2} + \frac{1-2\tau_3}{16}x^2 - \tau_0\right)\Phi(x) = 0 \quad (A.4)$$

where $\tau_0$ is the real eigenvalue which is equal to $2E$. Using only the commutation relations of SO(2,1) we can perform a unitary transformation called the "tilting transformation":

$$e^{i\zeta L_2}(L_3 \pm L_1)e^{-i\zeta L_2} = e^{\mp\zeta}(L_3 \pm L_1)$$

where $\zeta$ is a real constant parameter. This transforms Schrödinger equation (A.4) to

$$\left\{\left[1-(2\tau_3-1)e^{2\zeta}\right]L_1 + \left[1+(2\tau_3-1)e^{2\zeta}\right]L_3 - 2\tau_0 e^\zeta\right\}\Phi = 0$$

To obtain the discrete representation, we choose the "tilting angle" $\zeta$ such that the coefficient of $L_1$ vanishes. We should note, however, that the range of possible values of $\zeta$ is restricted by the value of $\tau_3$ in the problem. For bound states $\tau_3 > 1/2$, while, for scattering states (the continuum) $\tau_3 < 1/2$. However, presently we require that

$$(2\tau_3 - 1)e^{2\zeta} = 1 \quad \Rightarrow \quad \tau_3 > 1/2$$

Thus, giving

$$(L_3 - \tau_0 e^\zeta)\Phi = 0 \quad \Rightarrow \quad L_3\Phi = \frac{\tau_0}{\sqrt{2\tau_3 - 1}}\Phi$$

Using the spectrum of $L_3$ in equation (A.1), we can write

$$\frac{\tau_0}{\sqrt{2\tau_3 - 1}} = \gamma + n + 1 \quad ; \quad n = 0, 1, 2, ...$$

Now, if we define $(2\tau_3 - 1)/16 \equiv \lambda^4$, then we can finally write the differential equation (A.4) as follows

$$\left[\frac{d^2}{dx^2} - \frac{4\gamma(\gamma+1)+3/4}{x^2} - \lambda^4 x^2 + 4\lambda^2(\gamma+n+1)\right]\Phi_n^\gamma(x) = 0 \tag{A.5}$$

where $\Phi_n^\gamma(x) \equiv \langle x|\gamma,n\rangle$. The normalized solution of this differential equation can be written as [15]:

$$\Phi_n^\gamma(x) = \sqrt{\frac{2\lambda\,\Gamma(n+1)}{\Gamma(2\gamma+n+2)}}(\lambda x)^{2\gamma+3/2} e^{-\lambda^2 x^2/2} L_n^{2\gamma+1}(\lambda^2 x^2) \tag{A.6}$$

$\Gamma$ is the gamma function and $L_n^\nu(x)$ are the Laguerre polynomials. Equation (A.5) is, evidently, Schrödinger equation for the three-dimensional harmonic oscillator, where $\lambda$ is the oscillator strength parameter and $\gamma = (l - \tfrac{1}{2})/2$. It also gives the spectrum of the bound states whose corresponding normalized wavefunctions are those given by equation (A.6). PCTs map this differential equation into other solvable problems belonging to the same class as that of the oscillator. These transformations produce a correspondence map among the physical parameters of the two problems leading to the new spectrum and eigenstates wavefunctions. This is accomplished by performing coordinate transformations of equation (A.5) that preserves the Schrödinger–like structure of the equation (i.e., the first order derivative terms vanish). It should also leave the form of the function $f(x)$, as given by equation (A.3), invariant (shape-invariant). This transformation is written as

$$r = q(x) \quad, \quad \Psi(r) = g(x)\Phi(x)$$

The Schrödinger–like constraint requires that $g = \sqrt{dq/dx}$. Then, the differential equation reads

$$\left[\frac{d^2}{dr^2} + f(r)\right]\Psi(r) = (q')^{-3/2}\left\{\frac{d^2}{dx^2} + \frac{1}{2}\left[\frac{q'''}{q'} - \frac{3}{2}\left(\frac{q''}{q'}\right)^2\right] + (q')^2 f(q)\right\}\Phi = 0$$

where the prime symbol stands for the derivative with respect to the coordinate $x$. This equation is equivalent to (A.5) if we write:

$$f(r) = (q')^{-2}\left[-\frac{4\gamma(\gamma+1)+3/4}{x^2} - \lambda^4 x^2 + 4\lambda^2(\gamma+n+1) - \frac{1}{2}\frac{q'''}{q'} + \frac{3}{4}\left(\frac{q''}{q'}\right)^2\right]$$

$$\equiv -\frac{l(l+1)}{r^2} - 2V(r) + 2E$$

The choice $q(x) = x^2$ gives the exact solution (energy spectrum and eigenstate wavefunctions) of the nonrelativistic Coulomb problem. On the other hand, taking $q(x) = -\ln(x)$ solves the S-wave ($l = 0$) Mörse problem. As for the zero energy problems [11], the choice is $q(x) = x^{2\mu+1}$, where $\mu$ is a real parameter and $\mu \neq 0, \pm\tfrac{1}{2}$.